\documentclass[12pt]{article}
\usepackage[dvips]{graphicx}
\usepackage{epsfig,amsmath}

\begin{document}
\thispagestyle{empty}
\begin{center} 
{\bf HADRONIC AND SPIN PHYSICS: REVIEW \footnote{Invited talk at CIPANP 2009, San Diego California, USA, May 26-31, 2009, to appear in the AIP Conference Proceedings}}\\

\vskip 1.4cm
{\bf Jacques Soffer}
\vskip 0.3cm
Physics Department, Temple University, Philadelphia, PA 19122-6082, USA\\
\end{center}
\vskip 1.5cm

{\bf Abstract}
I will summarize the numerous contributions which were presented in the session
{\it Hadronic and Spin Physics}, largely dominated by new experimental results.

\section{Introduction}
Although some of the contributions to this session have been scheduled later in the week, after this review talk, I will try to
give you the most complete overview of all the topics presented with some discussions. Many new experimental results were shown, coming from RHIC at
BNL, JLab, COMPASS at CERN, HERMES at DESY together with future plans for several facilities. The subjects discussed fall into the following four
categories: Polarized parton distributions - Spin asymmetries in DIS and structure functions - Single transverse spin asymmetries:
Sivers, Collins - Azimuthal asymmetries in unpolarized DIS. 

\section{Polarized parton distributions}
The knowledge of polarized parton distributions is very important for a better understanding
of the proton spin structure and this has been undertaken, in the last few years, by using either polarized
Deep Inelastic Scattering (DIS), at various facilities, or polarized proton beams at
the Relativistic Heavy Ion Collider (RHIC) at BNL, a unique high energy polarized $pp$ collider. One of the aims
of the spin program at RHIC is the determination of the gluon spin contribution, $\Delta G$, to the proton spin
and this was achieved, at least at a preliminary stage, by both experiments PHENIX and STAR. $\Delta G(x)$ is
extracted from the double-longitudinal spin asymmetry $A_{LL}=(\sigma_{++} - \sigma_{+-})/(\sigma_{++} + \sigma_{+-})$, where
$++$ ($+-$) denotes same (opposite) sign of the helicity states of the two incoming proton beams. PHENIX measures $A_{LL}$ in
various single particle inclusive channels (e.g. $\pi^{0,\pm}$, $\gamma$, $e$ and $\mu$), which are sensitive to $\Delta G(x)$
and cover a wide range of $x$.
\begin{figure}[htb]
\hspace*{-0.5cm}
  \begin{minipage}{7.0cm}
  \epsfig{figure=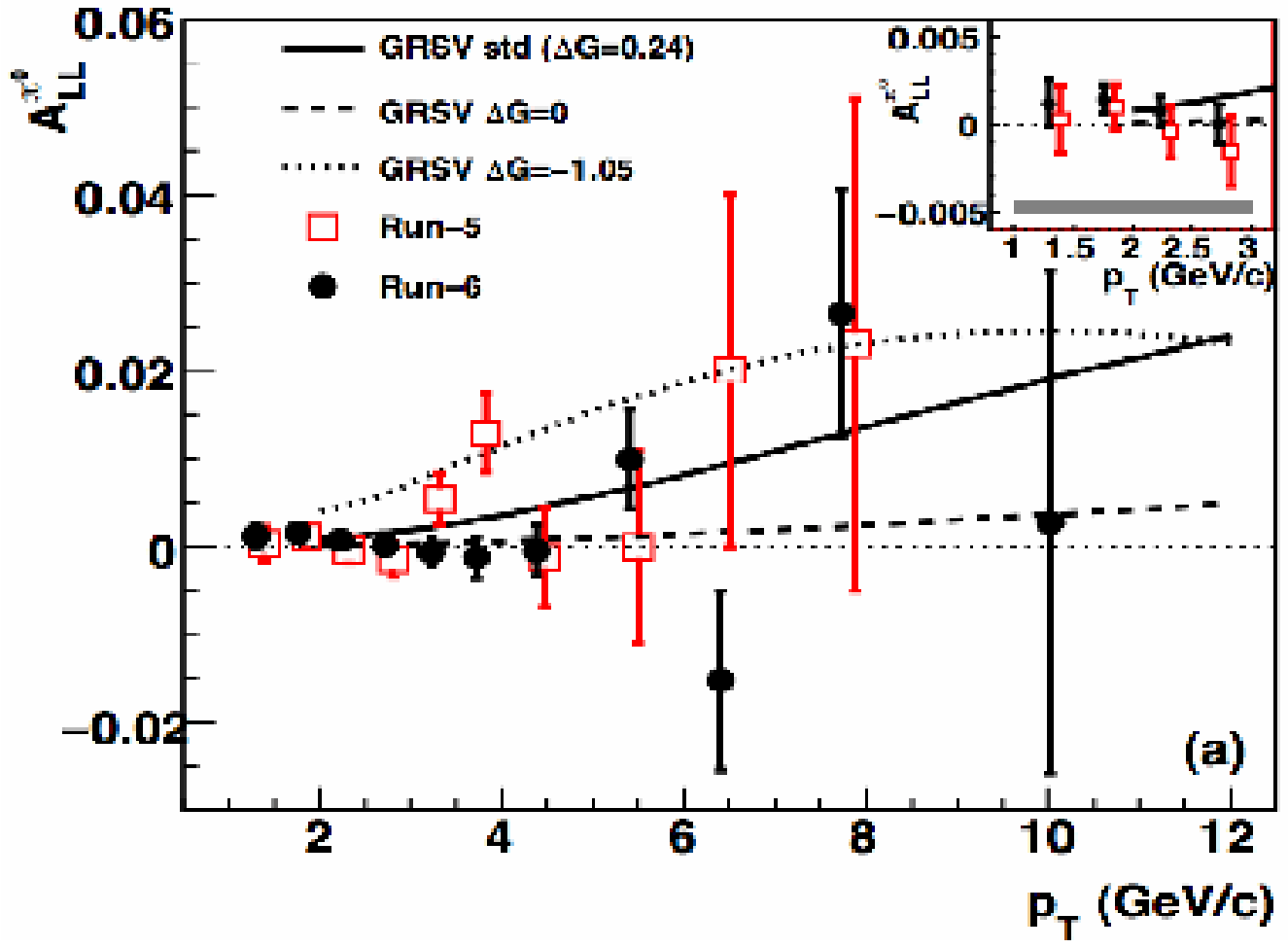,height=5.0cm}
  \end{minipage}
    \begin{minipage}{7.0cm}
  \epsfig{figure=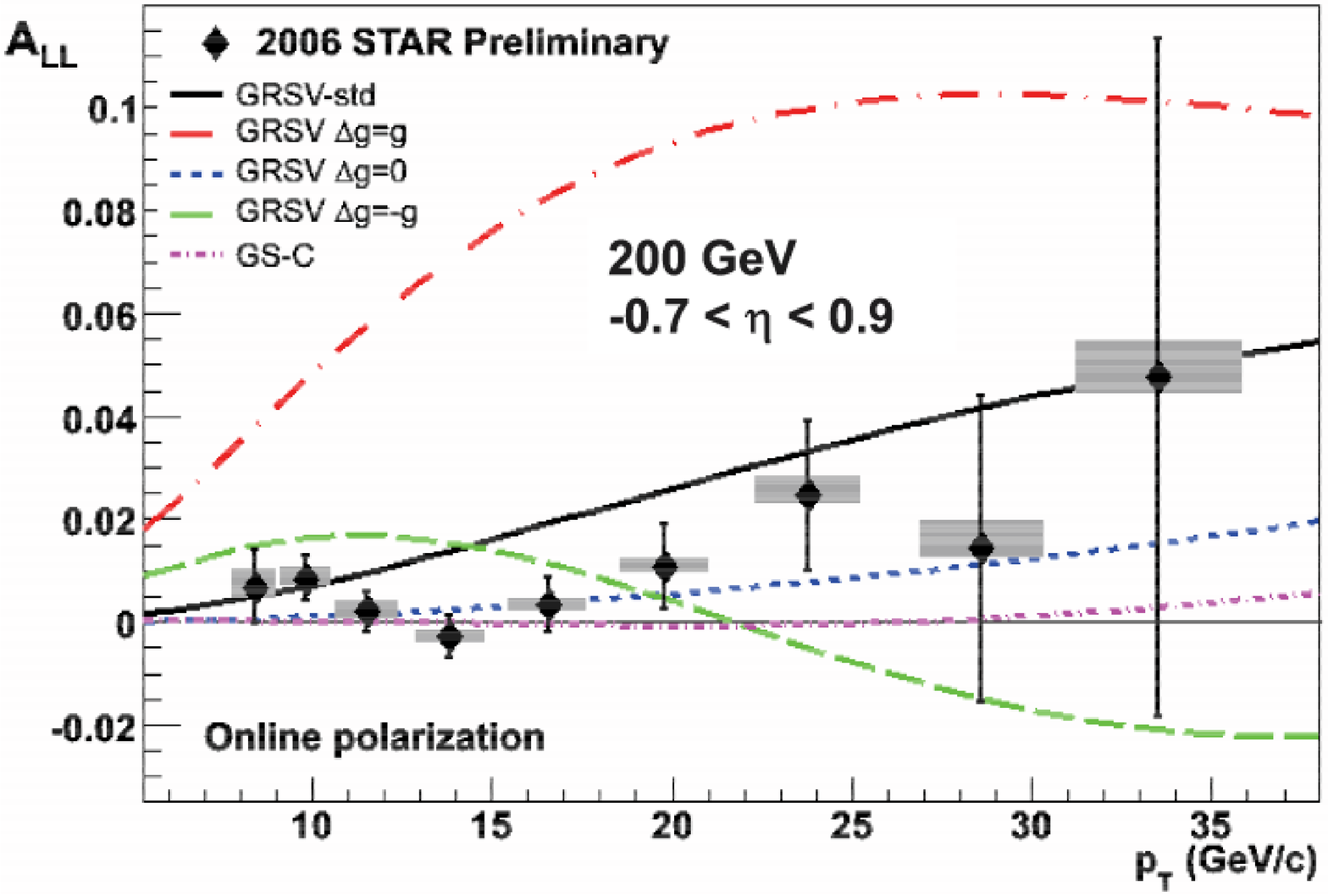,width=7.0cm}
  \vspace*{-2mm}
\end{minipage}\\  
\caption{Double-longitudinal spin asymmetries $A_{LL}$ from RHIC measurements at $\sqrt{s}=200\mbox{GeV}$ as a function of $p_T$ compared to several NLO pQCD theoretical expectations.
{\it Left}: PHENIX results for inclusive $\pi^0$ 
production (Taken from K. Barish talk). {\it Right}: STAR results for inclusive single-jet production (Taken from P. Djawotho talk).}
\label{fi:fig1}
\end{figure}
In the case of $\pi^0$ production, the results at $\sqrt{s}=200\mbox{GeV}$ are shown on Fig.~\ref{fi:fig1} and provide a
significant constraint on the integral of $\Delta G(x)$ in the $x$ range ~[0.02,0.3], namely $-0.7 < \Delta G ^{x=[0.02,0.3]}< 0.5$, at
the $3\sigma$ level \cite{phenix}. Other PHENIX data will be soon available and they will give access to the sign of $\Delta G(x)$ and
extend the $x$ coverage, which is crucial.
STAR has measured $A_{LL}$ in single-jet inclusive production with $p^{jet}_T$ up to 35 GeV/c and the latest results are shown on Fig.~\ref{fi:fig1}. They provide significant constraints and in particular they exclude large $\Delta G$.
Note that in both cases, PHENIX and STAR have checked the success of NLO pQCD cross sections calculations, which indicates that the
RHIC spin program has a solid theoretical ground.\\
Another important topic is the flavor separation of the various helicity distributions, specially for sea quarks, which can be done by using semi-inclusive polarized DIS, with identification of $\pi^{\pm}$ and $K^{\pm}$ in the final state. Some earlier results obtained by COMPASS, on a deuteron target, have been improved with a much higher statistics and the data cover the range $Q^2$ > ($1 \mbox{GeV/c}^2$) and $0.004<x<0.3$ (see
O. Kouznetsov talk). The distribution of $\Delta s(x)$ is compatible with zero in the whole measured range and the sensitivity upon the choice
of fragmentation functions was also discussed \cite{compass}.\\
It was pointed out long time ago \cite{bs} that the production of $W^{\pm}$ bosons in longitudinally polarized $pp$ collisions provides an
excellent tool to perform the flavor separation of $\Delta u(x)$, $\Delta d(x)$, $\Delta \bar{u}(x)$ and $\Delta \bar{d}(x)$. The essential
argument was recalled together with future plans of the STAR detector for the $W$ physics program at RHIC with $\sqrt{s}=500\mbox{GeV}$ 
(see W. Jacobs talk). The flavor breaking for the unpolarized light sea quarks has been established some time ago, {\it i.e.}  $\bar d(x) > \bar u(x)$,
and flavor breaking occurs also for the corresponding polarized distributions, since according to COMPASS results $\Delta \bar u(x) + \Delta \bar d(x)
\simeq 0$. All these facts remain to be confirmed at a much higher $Q^2$ value, so we expect exciting results from this $W$ physics program at RHIC, in the near future.
\section{Spin asymmetries in DIS and structure functions}
\begin{figure}[htb]
\hspace*{-0.5cm}
  \begin{minipage}{7.0cm}
  \epsfig{figure=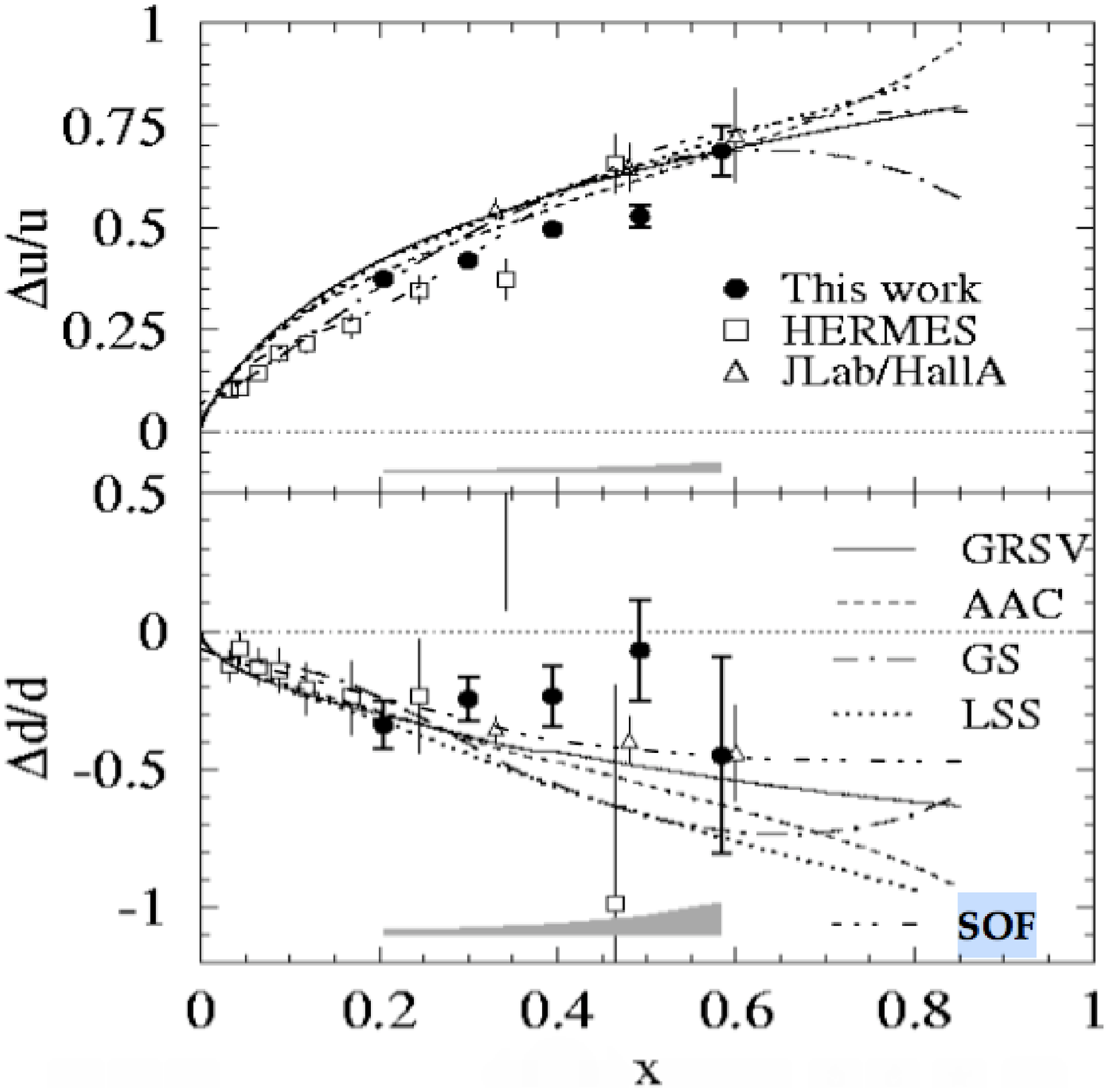,width=7.0cm}
  \end{minipage}
    \begin{minipage}{7.0cm}
  \epsfig{figure=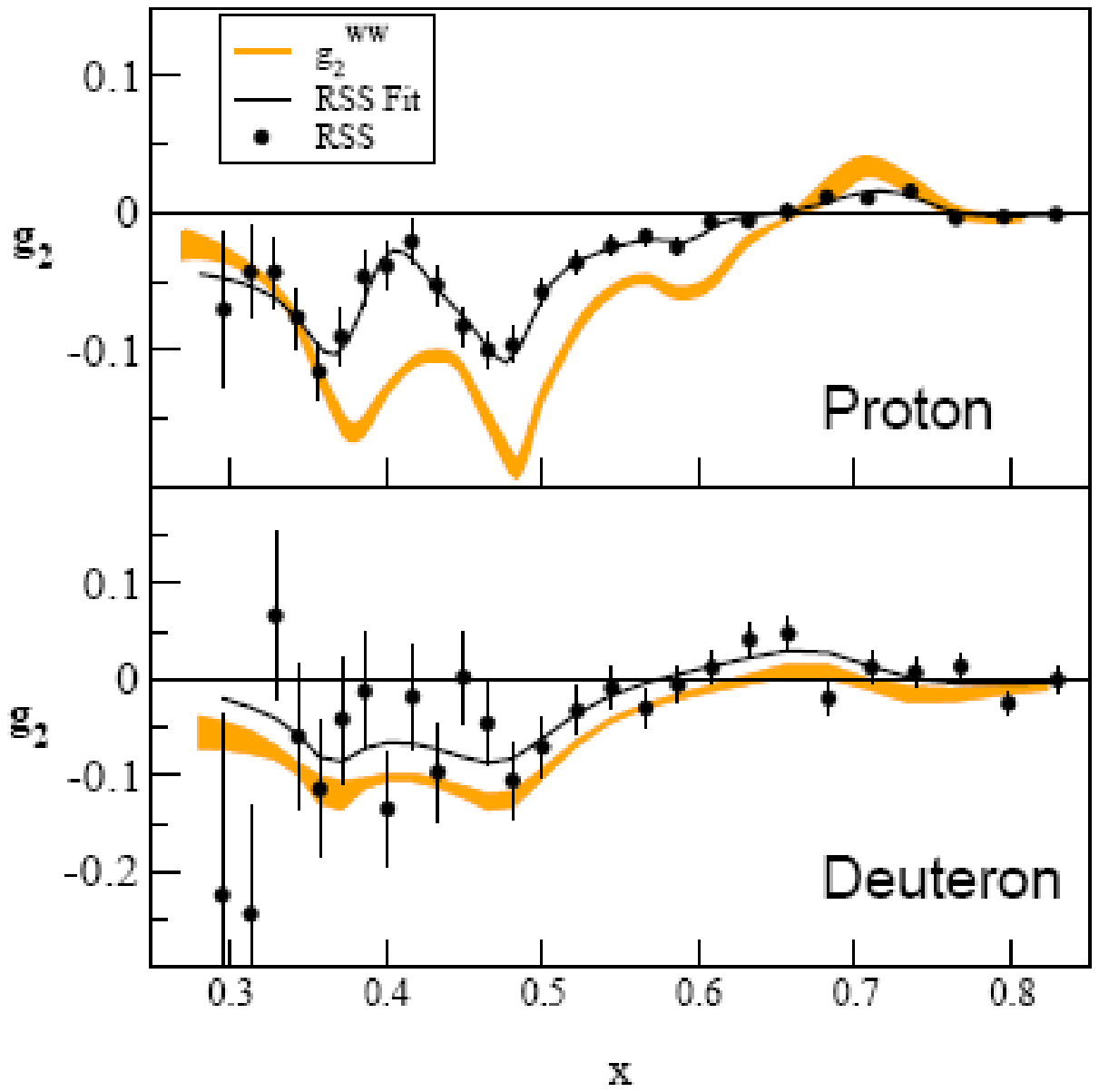,width=7.5cm}
  \vspace*{-3mm}
 \end{minipage}\\ 
\caption{
{\it Left}: Various data on quark polarization $\Delta u/u$ and $\Delta d/d$
versus $x$, compared to some theoretical expectations (Taken from Ref.~\cite{dhar}). The additional curve with the label SOF
is the prediction from Ref.~\cite{bbs}. {\it Right}: Proton and deuteron structure functions $g_2$ measured by the RSS experiment (Taken from M. Khandakar talk).}
\label{fi:fig2}
\end{figure}

Several experimental talks were devoted to different aspects of the usual proton and neutron polarized structure functions, namely
$g_1^{p,n}(x,Q^2)$ and $g_2^{p,n}(x,Q^2)$, extracted from the spin asymmetries 
$A_1$ and $A_2$ measured in polarized DIS at JLab, by using different polarized targets. The CLAS Collaboration has
reported, from the experiments EG1 and EG4 in Hall-B, preliminary data of impressive high precision for several $Q^2$ bins from 0.06 to 2.23 $\mbox{GeV}^2$ (see A. Biselli talk). The analyses of the full data will strongly constrain the phenomenological parametrizations of the
helicity distributions which can be extracted from these results. The left panel of Fig.~\ref{fi:fig2} shows a compilation of earlier published data compared
to theoretical expectations. Note that the behavior of $\Delta d(x)/d(x)$ in the large $x$ region is of crucial importance to discriminate 
between different predictions. Precise
data are also very usefull for sum rules studies involving the corresponding first moments of $g_1^{p,n}(x,Q^2)$, namely 
$\Gamma_1^{p,n}(Q^2)=\int_0^1 g_1^{p,n}(x,Q^2)dx$, which will be discussed later.\\
The RSS Collaboration in Hall-C has measured the proton and deuteron asymmetries in the resonance region and $Q^2 \simeq 1.3 \mbox{GeV}^2$
to search for the onset of local duality for $g_1$ and to explore the twist-three contribution to $g_2$. The right panel of Fig.~\ref{fi:fig2} displays
the RSS data on $g_2$ for protons and deuterons together with the twist-two components, showing a clear evidence for higher twist to $g_2^p$.\\

A bound on the DIS transverse asymmetry $A_2$ has been established
long time ago and reads $|A_2| \leq \sqrt{R}$, where
$R$ is the standard ratio $\sigma_L/\sigma_T$.
There is an improved version of this positivity constraint, namely,
$|A_2| \leq \sqrt{R(1+A_1)/2}$, which is very relevant when the DIS longitudinal asymmetry $A_1$ is negative. This
is the case for $A_1^n$, with a neutron target in a certain kinematic range. Several other significant positivity results for spin
observables based on Ref.~\cite{aerst}, were also presented (see J. Soffer talk).\\

\begin{figure}[htb]
\hspace*{-1.0cm}
\begin{minipage}{7.0cm}
  \epsfig{figure=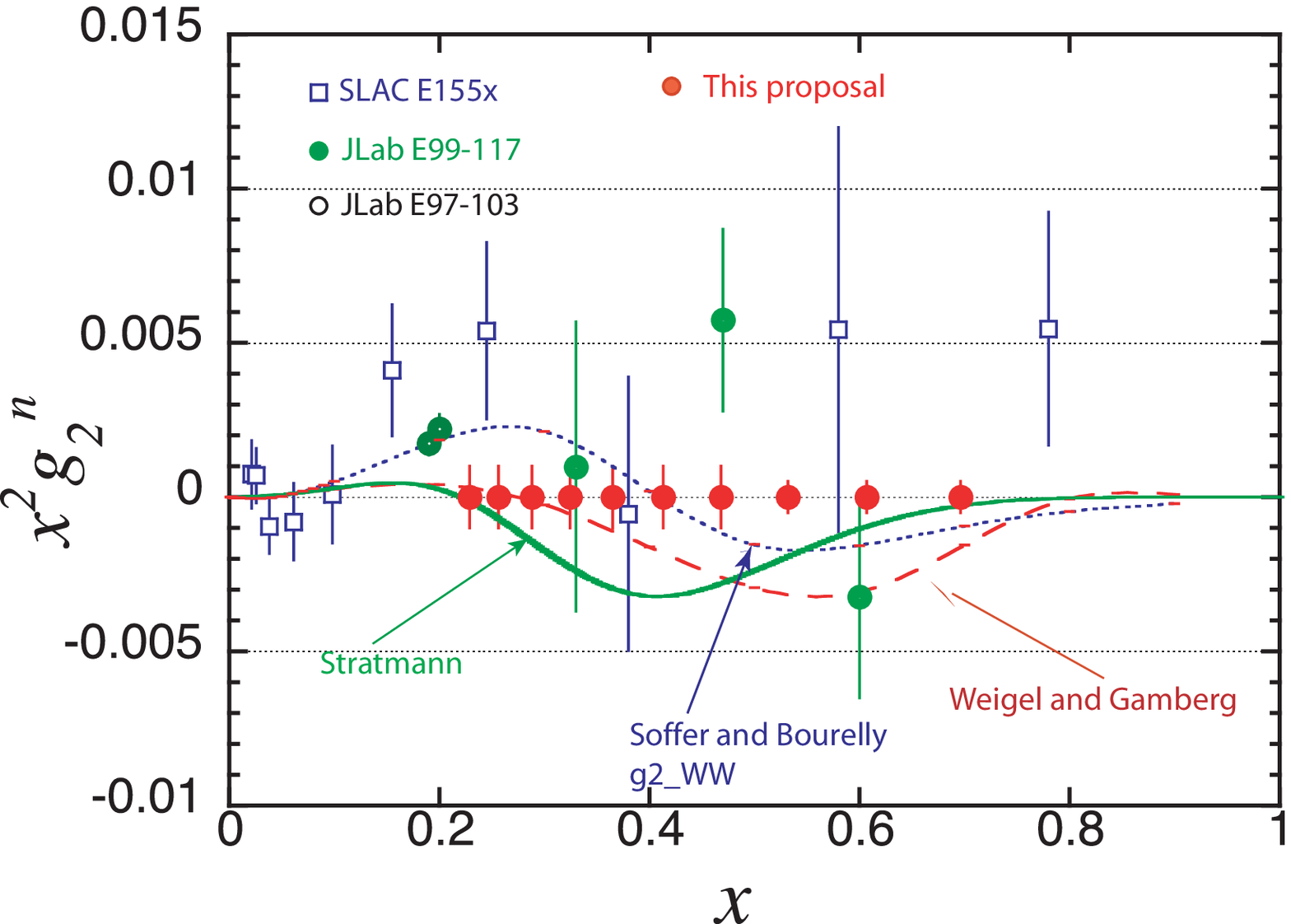,width=7.5cm}
  \end{minipage}
  \hspace*{1cm}
  \vspace*{-1.0ex}
    \begin{minipage}{7.0cm}
  \epsfig{figure=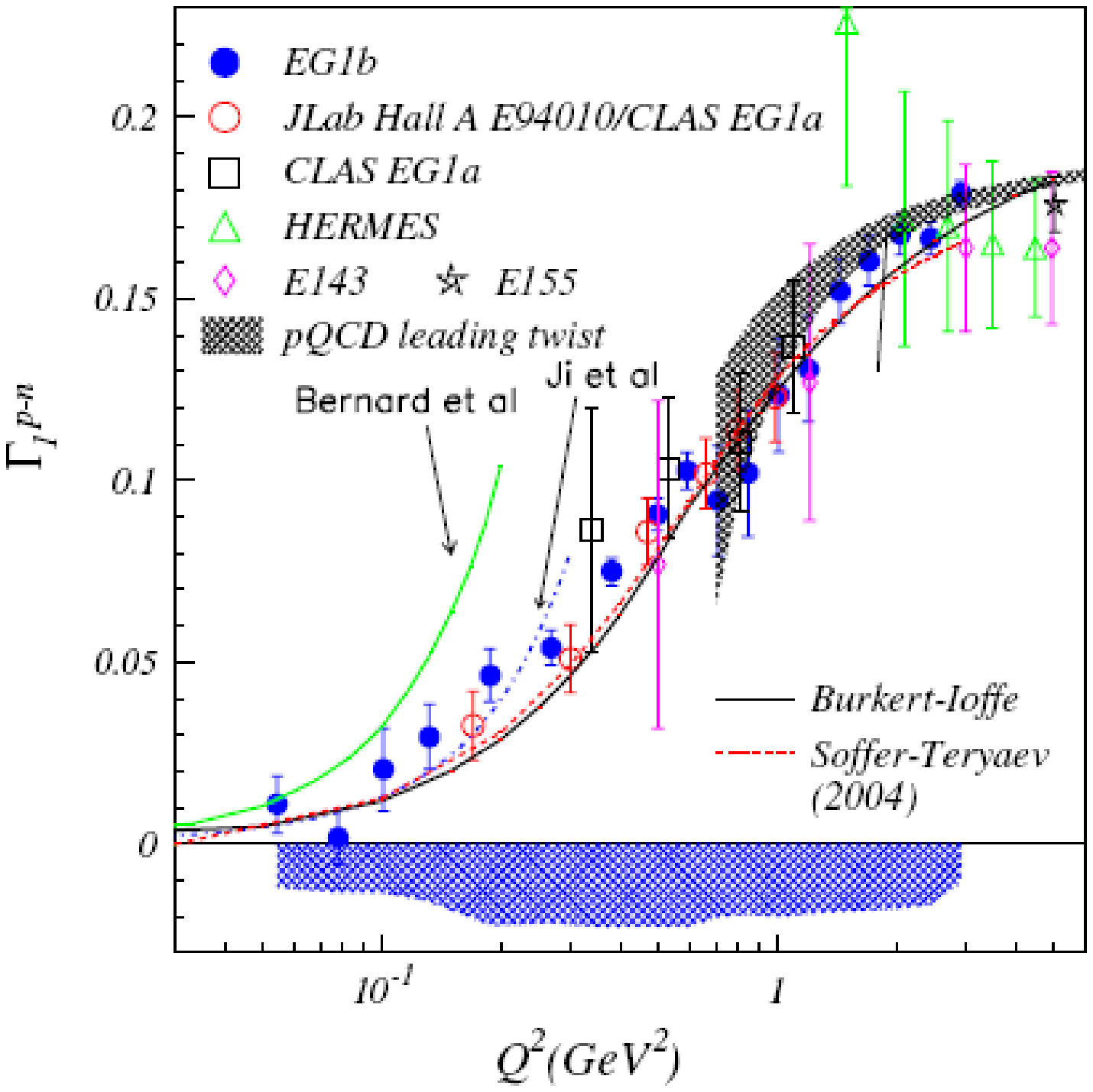,width=7.0cm}
  \vspace*{-3mm}
 \end{minipage}\\ 
\caption{
{\it Left}: World data on $g_2^n$ and expected precision of E06-114 with a polarized $^{3}$He target, compared to theoretical predictions (Taken from
Z.E. Meziani talk). {\it Right}: Bjorken integral $\Gamma_1^{p-n}(Q^2)$. The solid blue circles are from the JLab experiment EG1b for
$0.05 < Q^2 < 2.92 \mbox{GeV}^2$ (Taken from Ref.~\cite{deur}).}
\label{fi:fig3}
\end{figure}

Two recently ran experiments at JLab, E07-003 in Hall-C (proton) and E06-114 in Hall-A (neutron) were presented, whose aims are, in particular, to 
evaluate quark-gluon correlations. These higher twist contributions can be interpreted in terms of the concepts of "`color polarizabilities"' and average color Lorentz force (see Z.E. Meziani talk). Although the results will not be available before at least one year, we can already see on the left panel of Fig.~\ref{fi:fig3}, the expected improvement on the data in the neutron case. This program will be pursued at JLab upgrade to 12 GeV for higher precision and larger $x$ and $Q^2$ coverage.\\
There are many sum rules involving spin structure functions and, in particular, the Bjorken sum rule is known to be a fundamental test of
pQCD. Recent precision data from JLab, displayed on the right panel of Fig.~\ref{fi:fig3} show how its $Q^2$ behavior is well described by the theory and its smooth connection to the Gerasimov-Drell-Hearn sum rule. The subject was discussed extensively including the Burkhardt-Cottingham
sum rule (see J.P. Chen).

\section{ Single transverse spin asymmetries: Collins, Sivers}
For a long time the subject of single spin asymmetries (SSA) in high energy collisions was considered uninteresting
and irrelevant in the framework of the QCD parton model, because by using very naïve arguments, SSA were expected essentially to
be bound to zero. We recall that a nonzero SSA requieres an helicity flip amplitude having a phase difference with the nonflip amplitude. This was already in contradiction with the observation of a large SSA, nearly thirty years ago,
in $\pi$ inclusive production in $pp$ collisions at FNAL, with a transversely polarized proton beam. However according to some recent theoretical ideas, it is now possible to interpret sizeable SSA in terms of well defined nontrivial QCD dynamical mechanisms:\\
- The Sivers effect based on a correlation between the nucleon spin and the transverse momentum of the quark, sensitive to the quark orbital motion\\ 
- The Collins effect based on the quark transversity distribution and spin-orbit effects in the fragmentation\\
This leads to the consideration of new distributions, the Sivers and the Collins functions, which are transverse momentum dependent (TMD) and have
generated an extensive theoretical activity. For example, an important issue is the universality of the Collins function discussed in F. Yuan talk. He has identified the correspondent collinear twist-three fragmentation function for the Collins effect and showed that it is universal. On the contrary,
the Sivers function, which is directly connected to final state interaction has no universality. In particular, since final state interactions needed to
produce a SSA in DIS and the corresponding initial state interactions in Drell-Yan have different color interactions, we expect $SiversSSA|_{DY} = - SiversSSA|_{DIS}$. This opposite sign is of crucial importance to test a nontrivial QCD prediction and this is already planned at RHIC and other facilities. Finally, L. Gamberg in his talk was presenting some evaluations of the Sivers function related to the so-called "`chromodynamic lensing function"' and the generalized parton distribution $E$.\\
\begin{figure}[htb]
\hspace*{-1.0cm}
\begin{minipage}{7.0cm}
  \epsfig{figure=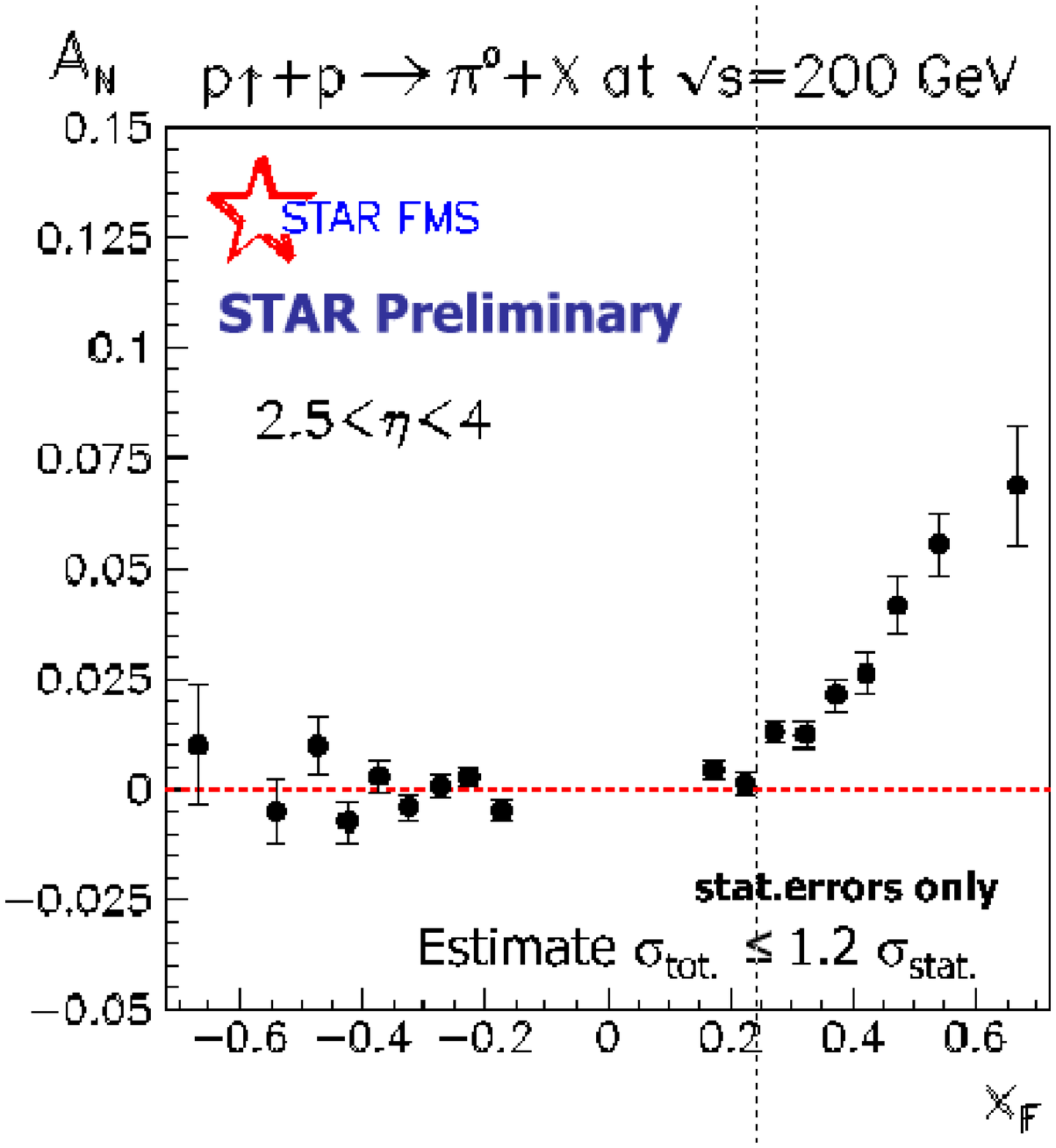,width=7.5cm}
  \end{minipage}
  \hspace*{0.5cm}
  \vspace*{-1.0ex}
    \begin{minipage}{7.0cm}
  \epsfig{figure=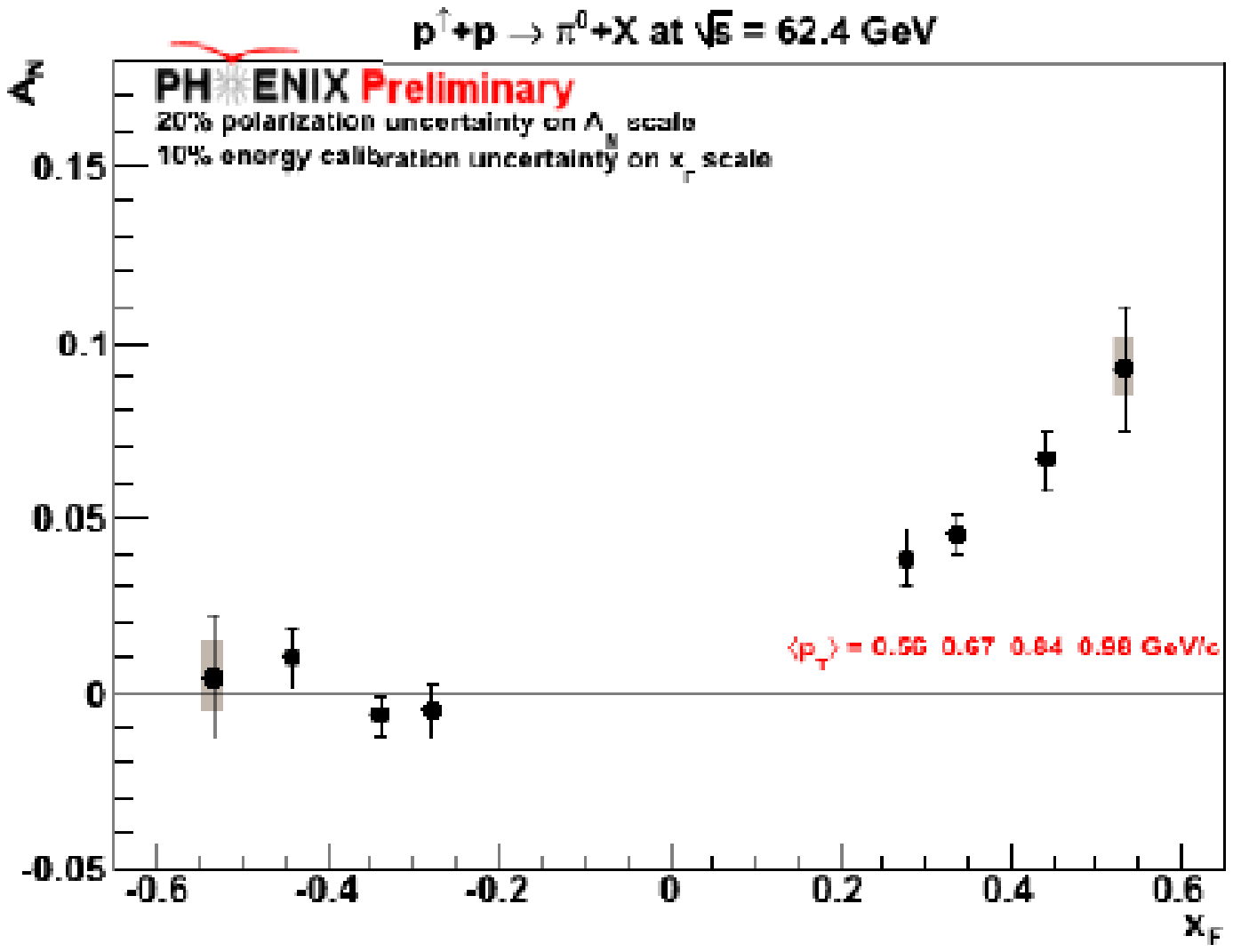,width=7.5cm}
  \vspace*{-3mm}
 \end{minipage}\\ 
\caption{
{\it Left}: SSA in $\pi^0$ inclusive production versus $x_F$ measured at STAR with $\sqrt{s}=200$ GeV (Taken from A. Ogawa talk). {\it Right}: SSA in $\pi^0$ inclusive production versus $x_F$ measured at PHENIX with $\sqrt{s}=62.4$ GeV (Taken from R. Yang talk).}
\label{fi:fig4}
\end{figure}
On the experimental side SSA in $\pi^0$ inclusive production have been measured at RHIC by STAR and PHENIX and the data are shown on Fig.~\ref{fi:fig4}. These results 
are compatible and seem to have almost no energy dependence. Moreover $A_N$ is positive and grows with $x_F$ for $x_F > 0$, whereas for $x_F < 0$, $A_N$ is
consistent with zero. In this case, since both mechanisms, Sivers and Collins, are allowed it is not easy to disentangle them. However this separation becomes possible in semi-inclusive DIS with a transversely polarized target, by measuring the hadron production on the sum and difference of two azimuthal angles with respect to the scattering plane. The procedure is explained in details and the results obtained by HERMES (see N. Makins talk) and by COMPASS (see R. Joosten talk) have been presented. By using these data combined with some results from BELLE on the fragmentation functions, it is possible to extract for the first time the transversity distributions $h_1^u(x)$ and $h_1^d(x)$.

\section{Azimuthal asymmetries in unpolarized DIS}
The semi-inclusive DIS cross section has an azimuthal dependence around the outgoing hadron direction in terms of cos $\phi_h$ and cos $2\phi_h$, where $\phi_h$ is the azimuthal angle of the hadron plane around the virtual-photon direction. Two mechanisms are expected to contribute to this
azimuthal dependence; the Cahn effect, a pure kinematic effect generated by the non-zero intrinsic transverse quark momentum \cite{cahn} and the
Bohr-Mulders effect resulting from a correlation between the transverse spin of a quark and its transverse momentum inside the hadron \cite{bm}. It is in fact the convolution of the Bohr-Mulders function with the Collins function which generates the effect. So
once more we see the importance of the TMD, which plays a central role also in this case. New HERMES results on <cos $\phi_h$> and <cos $2\phi_h$> with Hydrogen and Deuterium targets and positive, negative charged hadrons, were presented together with their interpretation from several theoretical models, knowing that the Boer-Mulders effect contributes to both terms (see R. Lamb talk). Similar new data from COMPASS were also presented which don't quite agree with those of HERMES (see R. Joosten talk), which might be a new challenge.\\

{\bf Acknowledgements} I would like to thank Douglas Beck for his help in preparing
the program of this session. I am very grateful to the organizers of the tenth Conference
on the Intersections of Particle and Nuclear Physics, CIPANP 2009, and in particular to Marvin
Marshak, for providing some financial support.

\end{document}